\documentclass[amsmath,amssymb,altaffilletter]{revtex4-1}
\usepackage{natbib}
\usepackage{amsmath}
\usepackage{graphicx}
\begin{document}

\title{Dispersion engineering of Quantum Cascade Lasers frequency combs}
\author{Gustavo Villares}
\email{gustavo.villares@phys.ethz.ch}
\author{Sabine Riedi}
\author{Johanna Wolf}
\author{Dmitry Kazakov}
\author{Martin J. S{\"u}ess}
\author{Mattias Beck}
\author{J\'er\^ome Faist}
\email{jfaist@phys.ethz.ch}
\affiliation{Institute for Quantum Electronics, ETH Zurich, Switzerland}

\begin{abstract} 
	
Quantum cascade lasers are compact sources capable of generating frequency combs. Yet key characteristics -– such as optical bandwidth and power-per-mode distribution -– have to be improved for better addressing spectroscopy applications. Group delay dispersion plays an important role in the comb formation. In this work, we demonstrate that a dispersion compensation scheme based on a Gires-Tournois Interferometer integrated into the QCL-comb dramatically improves the comb operation regime, preventing the formation of high-phase noise regimes previously observed. The continuous-wave output power of these combs is typically $>$ 100 mW with optical spectra centered at 1330 cm$^{-1}$ (7.52 $\mu$m) with $\sim$ 70 cm$^{-1}$ of optical bandwidth. Our findings demonstrate that QCL-combs are ideal sources for chip-based frequency comb spectroscopy systems. 
\end{abstract}

\maketitle
 
Optical frequency combs have revolutionized the fields of high-resolution and precision atomic spectroscopy due to their high coherence, wide spectral bandwidth and absolute traceability~\cite{udem_optical_2002,diddams2010evolving}. Initially developed in the near-infrared (NIR) spectral region, frequency combs are now being extended to other parts of the spectrum. In particular, extending the spectral range of frequency combs into the mid-infrared (MIR) and terahertz (THz) regions will open new possibilities in the fields of frequency metrology, molecular spectroscopy, chemical analysis and medical diagnosis~\cite{schliesser2012mid}, as the fundamental roto-vibrational absorption lines of a variety of molecules lie in this spectral region.
  
Different schemes have been investigated for generating MIR frequency combs. A well-established approach consists of transferring frequency combs from the near-infrared region into the MIR region through nonlinear processes using, for example, optical parametric oscillators~\cite{Adler:2009,Vodopyanov:2011} or difference frequency generation in fiber-based NIR combs~\cite{Ruehl:2012,Zhu:2013,Galli:2013d}. Other examples include MIR combs generated by transition metals incorporated into chalcogenide hosts~\cite{sorokin2012kerr,sorokin2012femtosecond} or Thulium-doped silica fiber lasers~\cite{solodyankin2008mode}. These sources are now well-established and applications such as MIR high-resolution spectroscopy are possible. These methods guarantee good spectral coverage and coherence, but usually require delicate experimental set-ups with large footprints. 

Significant effort has been recently made for achieving chip-based MIR frequency combs. Microresonator frequency combs (Kerr-combs) have been significantly improved~\cite{del2007optical,Herr:2012,herr2014temporal,chembo2013spatiotemporal} and have been extended to the MIR region~\cite{Wang:2013,Griffith:2015,lecaplain2015quantum}. Although Kerr-combs can be produced in different material platforms, they still require a high-power continuous wave (CW) laser as well as an evanescent coupling system, especially difficult to achieve in the MIR and THz regions.  

Quantum cascade lasers (QCL) have proven to be semiconductor lasers capable of generating comb radiation in the MIR and THz regions~\cite{hugi2012mid,Burghoff:2014,rosch2015octave}. As the comb formation takes place directly in the QCL active region, QCL frequency combs (QCL-combs) offer the unique possibility of a completely integrated chip-based system capable of performing broadband high-resolution spectroscopy. Such a compact system is ideal for applications requiring the detection of several different molecules masked by a complex background matrix.

Meanwhile, dual-comb spectroscopy using QCL-combs has been demonstrated~\cite{Villares:2014} and a theoretical description of the comb formation has recently been developed~\cite{Khurgin:2014,Villares:2015}. However, key characteristics of QCL-combs -– such as optical bandwidth and power-per-mode distribution -– still need to be improved in order to better address spectroscopy applications.

Group delay dispersion (GDD) plays an important role in the formation of QCL-combs~\cite{hugi2012mid,Burghoff:2014,Villares:2015}. In this work, we investigate a scheme for controlling the dispersion in MIR QCL-combs. We demonstrate that a dispersion compensation scheme based on a Gires-Tournois Interferometer~\cite{GTI_original} (GTI) directly integrated into the QCL-comb improves the comb performance. In particular, we show that the current range where the comb operates increases, effectively suppressing the high-phase noise regime usually observed in QCL-combs~\cite{hugi2012mid,Burghoff:2014,Villares:2014,burghoff2015evaluating,cappelli2015intrinsic}. Additionally, the power-per-mode distribution is improved. The CW output power of these combs is typically $>$ 100 mW and their optical spectra are centered at 1330 cm$^{-1}$ (7.52 $\mu$m) with $\sim$ 70 cm$^{-1}$ of optical bandwidth.

\subsection{Integrated Gires-Tournois interferometer for dispersion compensation}

Optical frequency combs are generated when the different longitudinal modes of a laser are locked in phase~\cite{udem_optical_2002,diddams2010evolving}, creating an array of equidistantly spaced phase-coherent modes. As demonstrated previously~\cite{hugi2012mid,friedli2013four}, broadband QCLs can achieve frequency comb operation by using four-wave-mixing (FWM) as a phase-locking mechanism. Combined with the short gain recovery time ($\tau \simeq 0.3$ ps) inherent of intersubband transitions, QCL-combs show a phase signature comparable to a frequency-modulated laser~\cite{hugi2012mid,Khurgin:2014}.

Efficient FWM process only occurs if the phase-mismatch $\Delta k$ between the modes involved in the FWM process nearly vanishes~\cite{boyd2003nonlinear,agrawal2007nonlinear}, \emph{i.e.}
\begin{align*}
	\Delta k = \frac{\tilde{n}_{4}\omega_{4} + \tilde{n}_{3}\omega_{3} -\tilde{n}_{2}\omega_{2} - \tilde{n}_{1}\omega_{1}}{c} \simeq 0
\end{align*}
where $\omega_{i}$ are the different mode frequencies involved in the FWM process and $\tilde{n}_{i}$ is the effective mode index at the frequency $\omega_{i}$. As the phase-mismatch condition depends on the effective mode indeces at different optical frequencies, a precise control of the dispersion of the laser is needed. More precisely, the phase-mismatch $\Delta k$ in a QCL can be expressed as
\begin{align*}
\Delta k = \Delta k_{\textrm{mat}} + \Delta k_{\textrm{wg}} + \Delta k_{\textrm{gain}} + \Delta k_{\textrm{NL}}
\end{align*}
where $\Delta k_{\textrm{mat}}$, $\Delta k_{\textrm{wg}}$ and $\Delta k_{\textrm{gain}}$ represent the phase-mismatch introduced by the material, by the laser waveguide and by the gain, respectively. The term $\Delta k_{\textrm{NL}}$ represents the phase-mismatch that can be introduced by self-phase and cross-phase modulation~\cite{agrawal2007nonlinear}. The advantage of QCLs regarding the generation of frequency combs is that these contributions can be tailored by design. As already shown experimentally in single-mode fibers~\cite{agrawal2007nonlinear,stolen1978self,hofer1991mode} and also in Kerr-combs~\cite{herr2014temporal,chembo2013spatiotemporal}, the FWM process starts to be efficient when working close to the zero-dispersion region. In this region, the different contributions $\Delta k_{\textrm{mat}}$, $\Delta k_{\textrm{wg}}$, $\Delta k_{\textrm{gain}}$ and $\Delta k_{\textrm{NL}}$, which may assume positive or negative values, start to be of comparable magnitude. One can therefore design one of the contributions to cancel the others and satisfy the phase-matching condition, thus enhancing the FWM process~\cite{lin1981phase}. The enhancement of the FWM efficiency is detrimental in the case of multichannel optical telecommunications systems, where high input powers introduce crosstalk between the different channels due to FWM~\cite{hui2002subcarrier}, but is an advantage in combs phase-locked by FWM~\cite{herr2014temporal,chembo2013spatiotemporal}.    

On this basis, we design the QCL-combs in the regime where the GDD is nearly zero. As QCLs are based on heterostructures where the composition can be tailored, the material dispersion can be controlled. The MIR QCL-combs used in this study are based on In$_{0.60}$Ga$_{0.40}$As/In$_{0.355}$Al$_{0.665}$As heterostructures grown on InP. Also, we optimize the mode profile in the waveguide for reducing the contribution of waveguide dispersion. This optimization is achieved by adjusting the number of periods of the QC structure and the doping profile of the InP cladding layer grown on top of the active region. Finally, the laser gain design is based on 2 different bound-to-continuum strain-balanced designs in the active regions, which are designed in order to minimize the dispersion introduced by the gain~\cite{hugi2012mid}. 
\begin{figure*}[t!]
	\centering
	\includegraphics[width=0.8\textwidth]{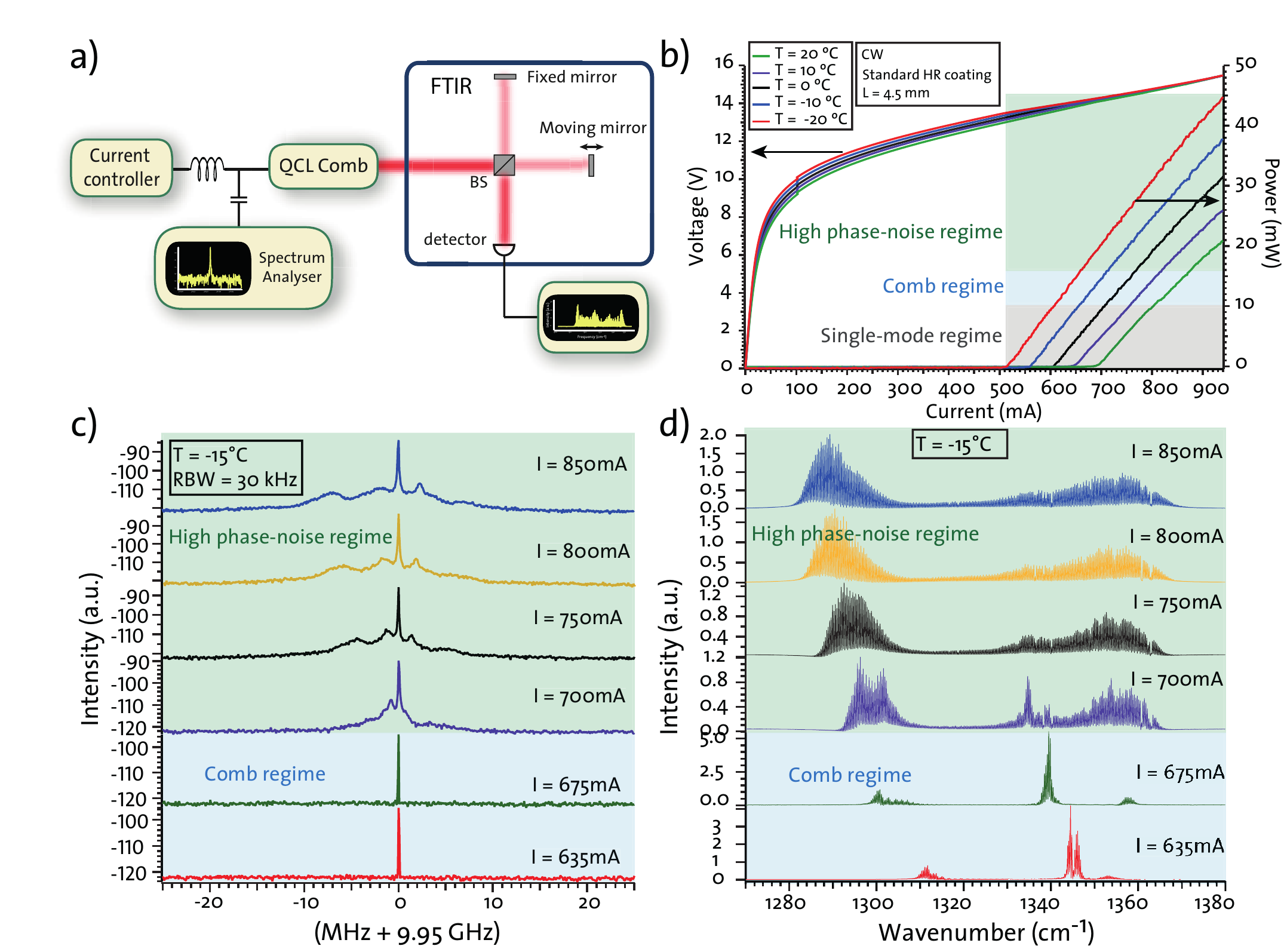}
	\caption{
		Standard QCL-comb performances.
		\textbf{a} Set-up used for characterizing the QCL-comb. The optical spectrum is measured with a FTIR (Bruker IFS 66/S, 0.12 cm$^{-1}$ resolution). A bias-tee is inserted between the low-noise current driver (Wavelength Electronics) and the QCL-comb. The radio-frequency (RF) port of the bias-tee is connected to a Spectrum Analyser (Rhode \& Schwarz FSU50). BS.: beam-splitter. FTIR: Fourier Transform Infrared Spectrometer.
		\textbf{b} Power-current-voltage of a QCL-comb (4.5 mm long, standard HR coating on the back facet, episide-down mounted on AlN submount) in CW operation at different temperatures. Single-mode, comb and high-phase noise regimes are highlighted.
		\textbf{c} Electrical RF spectra acquired at T = -15 $^{\circ}$C at different values of current, measured with a spectrum analyser (span = 50 MHz, resolution bandwidth (RBW) = 30 kHz, acquisition time = 20 ms). The RF spectra are centered at 9.95 GHz, corresponding to the RF beatnote created by a 4.5 mm long device. Comb and high-phase noise regimes are highlighted.
		\textbf{d} Optical spectra acquired at T = -15 $^{\circ}$C at the same values of current as in Fig.\,\ref{fig:QCL_comb_starting_point}\textbf{c} and measured with a FTIR (0.12 cm$^{-1}$ resolution). The QCL-comb spectrum is centered at 1325 cm$^{-1}$ and spans over 60 cm$^{-1}$ in the comb regime.
	}
	\label{fig:QCL_comb_starting_point}
\end{figure*} 

Fig.\,\ref{fig:QCL_comb_starting_point} shows the typical performance of MIR QCL-combs engineered in order to operate near the zero-dispersion region. The comb optical spectra and repetition frequency are measured simultaneously (c.f. Fig.\,\ref{fig:QCL_comb_starting_point}\textbf{a} and methods). The device operates at room-temperature emitting $>10$ mW of output power in CW operation (c.f. Fig.\,\ref{fig:QCL_comb_starting_point}\textbf{b}). More importantly, we also report in Fig.\,\ref{fig:QCL_comb_starting_point}\textbf{b} the three different regimes that are typically observed in QCL-combs~\cite{hugi2012mid,Burghoff:2014,Villares:2014,burghoff2015evaluating,cappelli2015intrinsic}. The laser emits single-mode radiation after the laser threshold. After a second threshold, the laser operates in a \emph{comb regime}. Finally, at higher values of current, we observe a third regime, called \emph{high phase-noise regime} hereafter. These three different regimes are well observed when measuring the radio-frequency (RF) beatnote and the optical spectrum, as shown in Fig.\,\ref{fig:QCL_comb_starting_point}\textbf{c} and Fig.\,\ref{fig:QCL_comb_starting_point}\textbf{d}, respectively. In the single-mode regime, no RF beatnote is observed. The comb regime is characterized by a single low-noise beatnote, corresponding to a regime where all the modes are phase-locked~\cite{hugi2012mid} and equidistantly spaced~\cite{Villares:2014}. Finally, the high-phase noise regime is identified by a broader beat note. In this high-phase noise noise regime, both amplitude and phase noise of the laser are significantly higher then in the comb regime~\cite{cappelli2015intrinsic}. Comb operation is therefore achieved only for a narrow range of the dynamic range of the laser operation. Moreover, comb operation is observed in regions relatively close to laser threshold, where both output power and optical spectral bandwidth are small compared to roll-over, as shown in Fig.\,\ref{fig:QCL_comb_starting_point}\textbf{b}. Finally, we also observe in Fig.\,\ref{fig:QCL_comb_starting_point}\textbf{d} that the power distribution between the modes is highly inhomogeneous, which is not optimal for spectroscopy applications based on frequency combs~\cite{Villares:2014}.

 \begin{figure*}[t!]
 	\centering
 	\includegraphics[width=0.75\textwidth]{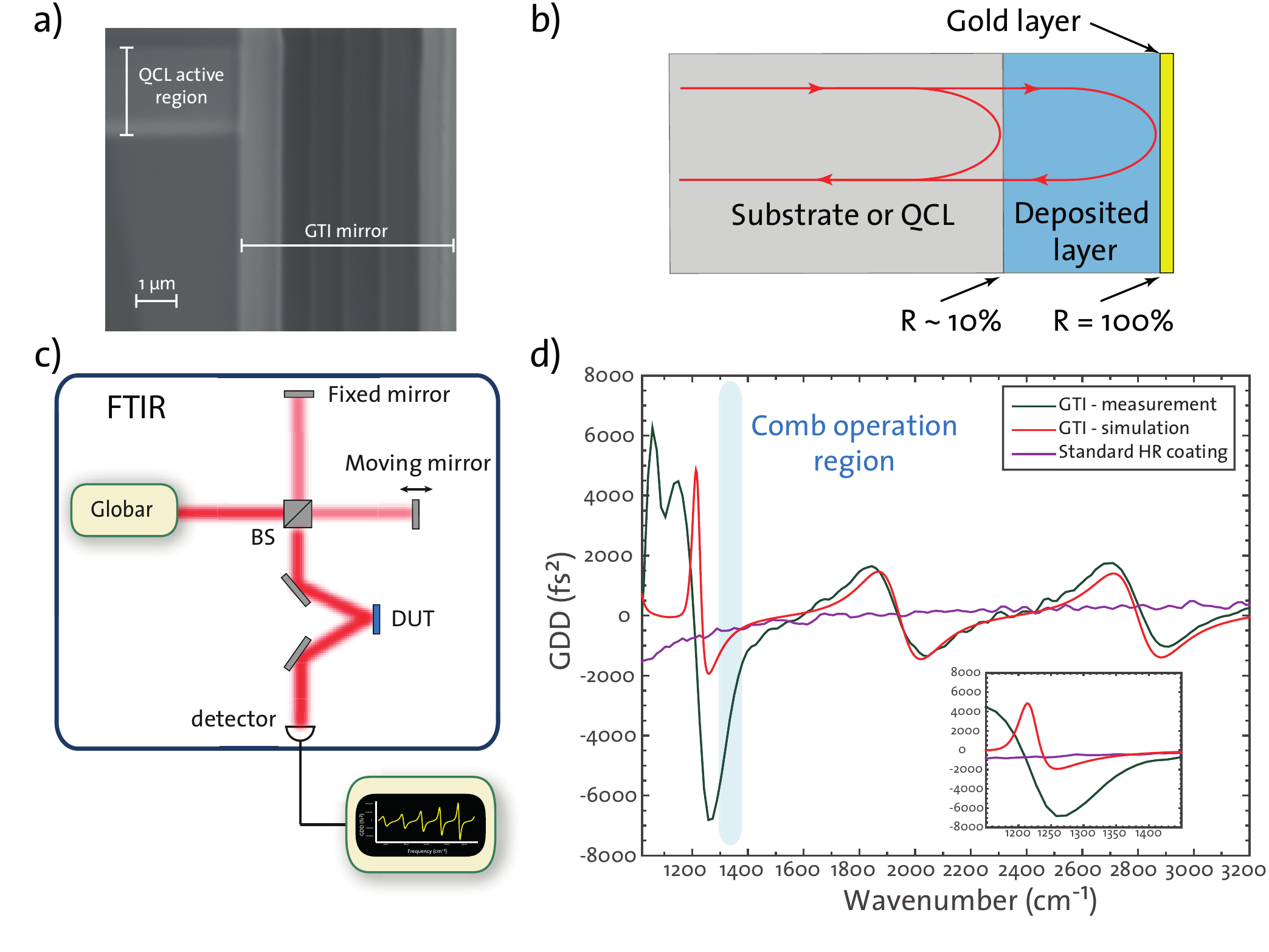}
 	\caption{
 		GTI mirrors for dispersion compensation. 
 		\textbf{a} SEM picture of a cross section parallel to the laser ridge of the QCL-comb, which is coated with a GTI mirror. The upper left side shows the laser active region. The different layers of the GTI mirror can be observed as the vertical lines on the right side of the picture.
 		\textbf{b} Schematic view of  GTI mirror coated either on the back-facet of a QCL-comb or on a substrate (InP, 320 $\mu m$ thick) to be used for dispersion characterization. The GTI acts as a high-relectivity mirror but adds a frequency dependent group delay, therefore introducing dispersion.
 		\textbf{c} Set-up used for the characterization of the dispersion introduced by the GTI mirror. The GTI mirrors coated on a substrate are measured in reflection on the sample compartment of the FTIR (see Appendix~\ref{App:disp_char}). DUT: device under test.  
 		\textbf{d} Measured and simulated value of the GDD created by a GTI mirror. The GDD is measured over a wide spectral range in order to observe the GDD oscillations introduced by the GTI. The spectral region where the QCL-comb operates is highlighted. The GDD of a standard HR coating (300 nm of $\textrm{Al}_{2}\textrm{O}_{3}$, 150 nm of gold) is also represented. Inset: Zoom on the spectral region where the QCL-comb operates, showing the negative GDD introduced due to the presence of the residual absorption of $\textrm{SiO}_{2}$. 
 	}
 	\label{fig:GTI}
 \end{figure*}

In order to further control the dispersion of QCL-combs, we integrate a GTI mirror~\cite{GTI_original} on the back-facet of the QCL-comb. Extensively used in solid state based mode-locked lasers \cite{szipocs2000negative}, GTI mirrors are optical cavities specifically designed for introducing dispersion. Fig.\,\ref{fig:GTI}\textbf{a} shows a cross-section of a QCL-comb coated with a GTI mirror taken with a scanning electron microscope (SEM) and Fig.\,\ref{fig:GTI}\textbf{b} shows a schematic of the integration of a GTI mirror on a QCL-comb. The GTI mirror is directly deposited on the back-facet of the device and is composed of several layers of $\textrm{Al}_{2}\textrm{O}_{3}$ and $\textrm{SiO}_{2}$ and terminated with a gold layer (see methods and Appendix~\ref{App:disp_char}). Assuming no absorption is present on the coating, a GTI mirror usually constitutes a broadband high-reflectivity (HR) coating. In addition, dispersion is introduced as the phase of the reflected light becomes frequency dependent due to the resonance effect introduced by the optical cavity. The dispersion introduced by a GTI is periodic with a period dependent on the length and on the refractive index of the material. By careful control of these parameters, a GTI mirror can introduce positive or negative dispersion to the QCL-comb (see Appendix~\ref{App:disp_char}). 
   
Different types of GTIs were evaporated on the back-facet of several devices. During each evaporation, a InP substrate was coated  as a reference sample so the dispersion introduced by the GTI mirror can be characterized independently of the device on which it was evaporated. A FTIR was used to measure the complex reflection spectrum of the coating (see Fig.\,\ref{fig:GTI}\textbf{c} and Appendix~\ref{App:disp_char}). Fig.\,\ref{fig:GTI}\textbf{d} shows the measured values of the GDD introduced by the GTI mirror, as well as the GDD obtained by simulation (see methods). The typical periodic variations of the GDD of a GTI are observed. Whenever the introduction of negative-dispersion is desired, the GTI can be designed such that one of its negative resonances lies in the spectral region of the respective QCL-comb. This is depicted on Fig.\,\ref{fig:GTI}\textbf{d} where we highlight the part of the spectrum where the comb is operating. At this particular resonance, situated around 1300 cm$^{-1}$, we observe a disagreement between the simulated value of GDD and its measured value. This disagreement is attributed to the fact that $\textrm{SiO}_{2}$ starts to be absorbing in this spectral region (see Appendix~\ref{App:disp_char}). This absorption adds a contribution to the total GDD introduced by the GTI. The value introduced at the minimum of this resonance is measured to be $\simeq$ -7000 fs $^{2}$ (see inset of Fig.\,\ref{fig:GTI}\textbf{d}). We also use the same measurement technique for characterizing the dispersion introduced by a standard HR coating (300 nm of $\textrm{Al}_{2}\textrm{O}_{3}$, 150 nm of gold). As expected, the GDD of a standard HR coating does not show any resonance effect and does not add any significant dispersion to the device. 

\subsection{Dispersion measurements in QCL-combs}

\begin{figure*}[t!]
	\centering
	\includegraphics[width=0.7\textwidth]{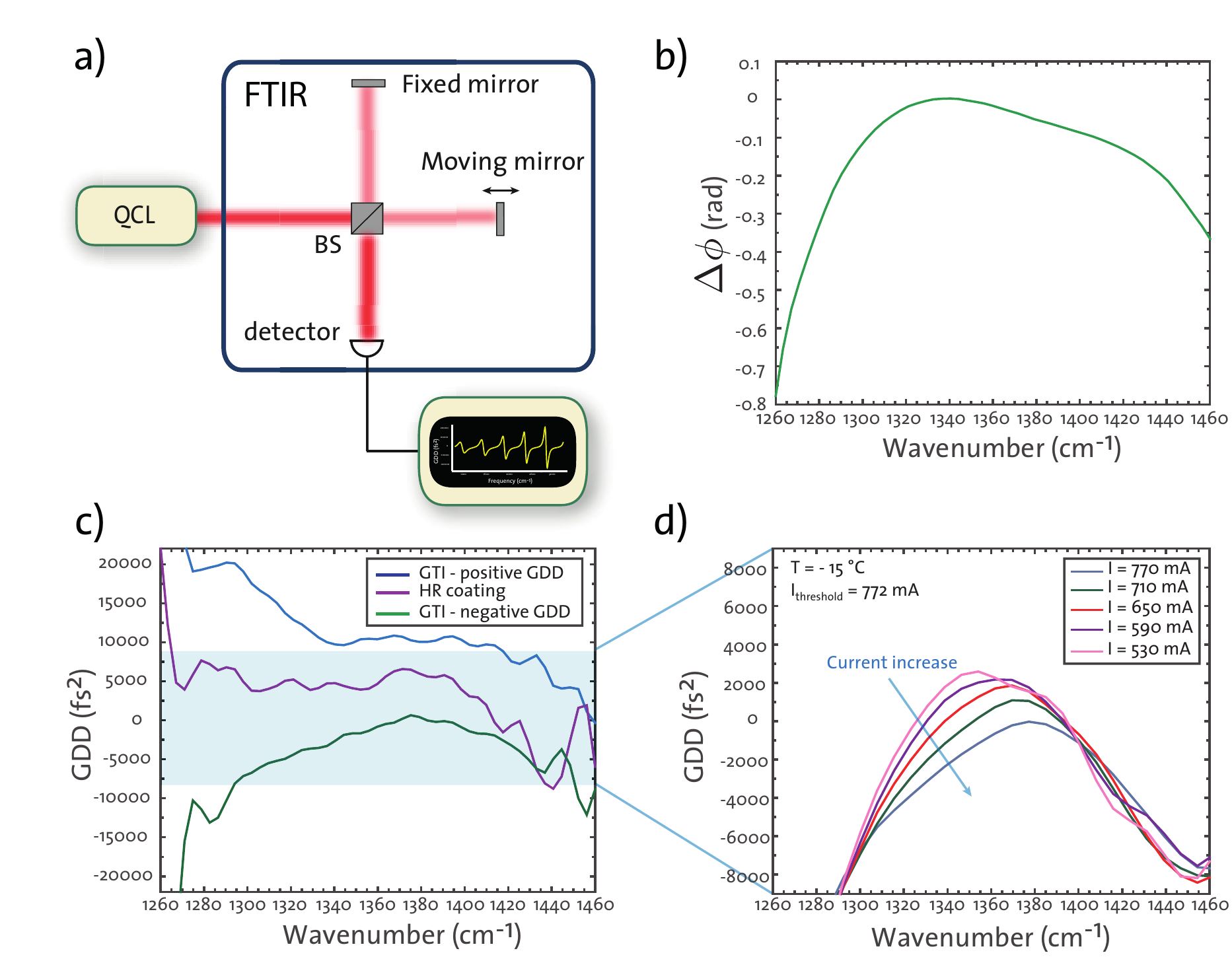}
	\caption{
		Dispersion measurements of QCL-combs. 
		\textbf{a} Set-up used to acquire the interferogram generated by the QCL-comb biased below threshold on a FTIR. This interferogram is used to retrieve the relative phase accumulated through a round-trip on the device (see Appendix~\ref{App:disp_char}). 
		\textbf{b} Relative phase accumulated through a round-trip on the device for a QCL-comb coated with a GTI mirror introducing negative dispersion (T = -15 $^{\circ}$C).
		\textbf{c} Measurement of the GDD of QCL-combs (T = -15 $^{\circ}$ C). Three different coatings were evaporated on the back-facet of three different devices (4.5 mm long devices cleaved together).
		\textbf{d} Measurement of the GDD of QCL-comb as a function of the laser current (T = -15 $^{\circ}$C).  
	}
	\label{fig:dispersion_QCL}
\end{figure*}
For further evaluating the dispersion compensation technique, we also measure the dispersion of QCL-combs after being coated with a GTI mirror. GTI mirrors introducing positive and negative dispersion were evaporated on different QCL-combs. Particular attention was given to use close to identical devices by using lasers with the same dimensions (ridge width and length) and from the same fabrication process. On the previous section, a QCL-comb coated with a standard HR coating was characterized (see Fig.\,\ref{fig:QCL_comb_starting_point}) and we use this device as our reference sample.

\begin{figure}[t!]
	\centering
	\includegraphics[width=0.70\textwidth]{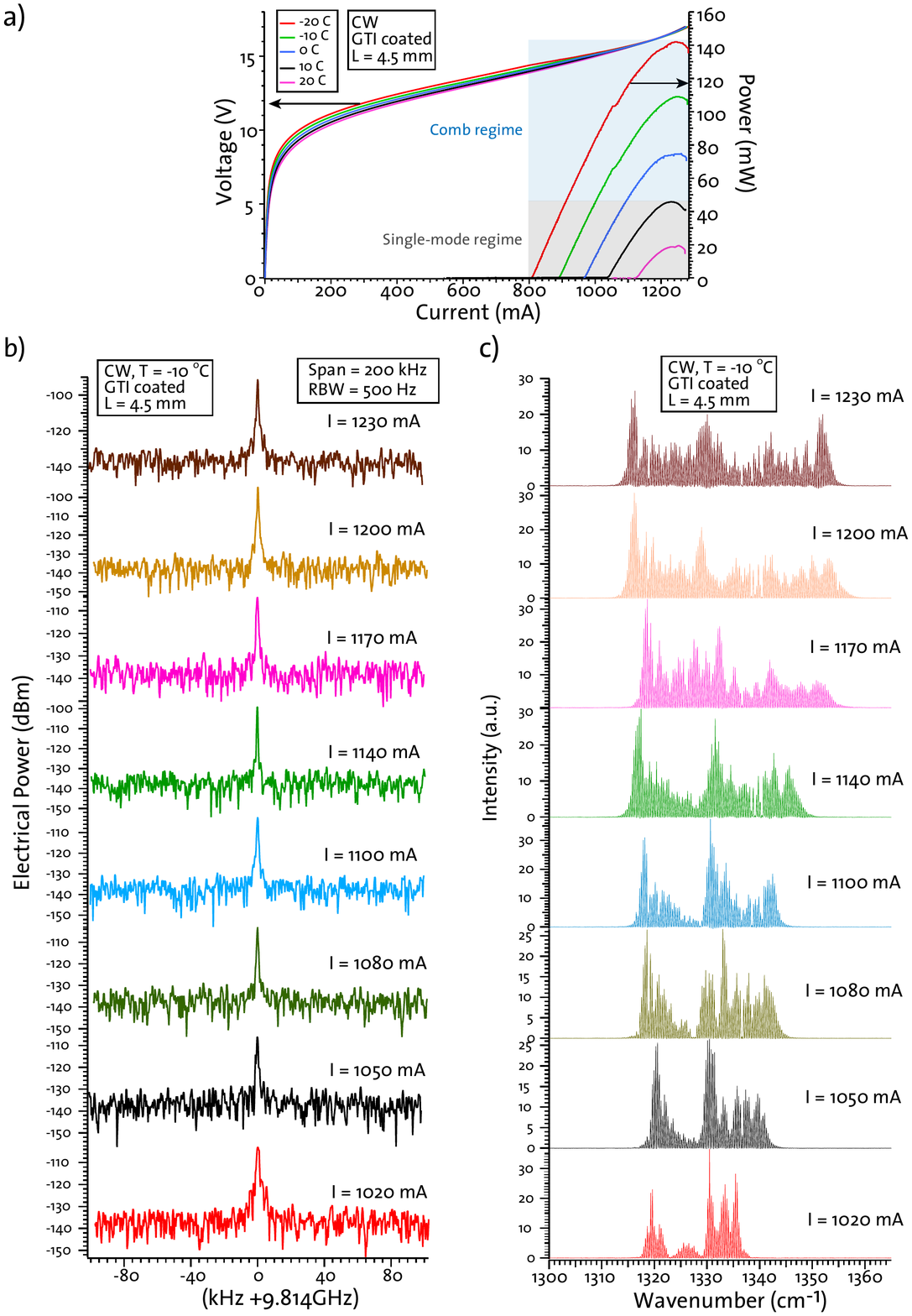}
	\caption{
		Dispersion compensated QCL-combs.
		\textbf{a} Power-current-voltage of a QCL-comb (4.5 mm long, episide-down mounted on AlN submount) coated with a GTI mirror introducing negative dispersion. The measurements are done in CW operation at different temperatures. Single-mode and comb regimes are highlighted. 
		\textbf{b} Electrical RF spectra acquired at T = -10 $^{\circ}$C for different values of current, measured with a spectrum analyser (span = 200 kHz, RBW = 500 Hz, acquisition time = 20 ms). The RF spectra are centered at 9.814 GHz, corresponding to the RF beatnote created by a 4.5 mm long device. The measured RF spectra show single and narrow beatnotes (FWHM $<$ 500 Hz). No high-phase noise regime is observed. 
		\textbf{c} Optical spectra acquired at T = -10 $^{\circ}$C at the same values of current as in Fig.\,\ref{fig:High_perf_QCL_comb}\textbf{b} and acquired with a FTIR (0.12 cm$^{-1}$ resolution). The QCL-comb spectrum is centered at 1335 cm$^{-1}$ and spans over 45 cm$^{-1}$ in the comb regime. 	
	}
	\label{fig:High_perf_QCL_comb}
\end{figure}
The dispersion of QCL-combs is measured by driving the QCL-comb close to but below threshold and acquiring the interferogram generated by a FTIR, as shown schematically in Fig.\,\ref{fig:dispersion_QCL}\textbf{a}. By careful analysis of the interferogram, the relative phase accumulated through a round-trip on the device can be extracted and the GDD of the QCL-comb can therefore be measured (see Appendix~\ref{App:disp_char}). Fig.\,\ref{fig:dispersion_QCL}\textbf{b} shows the relative phase of a device coated with a GTI mirror introducing negative dispersion, when the device is biased 2\% below threshold. The dispersion of devices coated with different GTI mirrors is shown in Fig.\,\ref{fig:dispersion_QCL}\textbf{c}. The determination of the dispersion is limited to the spectral range of the active region gain bandwidth (typically from $1250$ cm$^{-1}$ to $1460$ cm$^{-1}$), as this method is based on the subthreshold measurements (see Appendix~\ref{App:disp_char}). The QCL-comb coated with a standard HR coating is operating with a total positive GDD of $4131$ fs$^{2}$ (measured at $1330$ cm$^{-1}$). A similar device coated with a GTI mirror introducing positive GDD shows a total dispersion of $10602$ fs$^{2}$. The difference between the values of GDD of these two devices correspond to the measured value of the GDD measured in the reference GTI mirror, thus substantiating the claim that the added GDD is due to the engineered GTI coating. Finally, we also measure the dispersion of the device coated with a GTI mirror designed to introduce negative dispersion. This device is operating with a total negative GDD of $-3546$ fs$^{2}$.

In order to justify the introduction of the term $\Delta k_{\textrm{gain}}$ to the
phase-matching condition for QCL-combs, the effect of gain on the dispersion was investigated by measuring the GDD of a QCL-comb as a function of the laser current. The device is always driven below threshold, as the intention is to study the dispersion of the device with no effect of gain clamping. Fig.\,\ref{fig:dispersion_QCL}\textbf{d} shows the GDD of the device coated with a GTI mirror introducing negative dispersion for different driving currents (subthreshold measurements). The peak GDD value changes from $2591$ fs$^{2}$ at I $= 530$ mA to $- 26$ fs$^{2}$ at I $ = 770$ mA, demonstrating that the effect of gain on the total device dispersion is significant. Therefore, this measurement demonstrates that the gain-induced dispersion has to be considered when designing QCL-combs.      

\subsection{Suppression of the high-phase noise regime}

We now characterize the performances of QCL-combs where the dispersion was controlled by GTI mirrors. As shown in Fig.\,\ref{fig:dispersion_QCL}\textbf{c}, devices coated with a standard HR coating show small but positive dispersion values. Moreover, devices coated with GTI mirrors introducing positive dispersion showed the same performances as QCL-combs coated with standard HR coatings. Therefore, only the devices coated with GTI mirrors introducing negative dispersion are investigated here.
\begin{figure}[t!]
	\centering
	\includegraphics[width=0.8\textwidth]{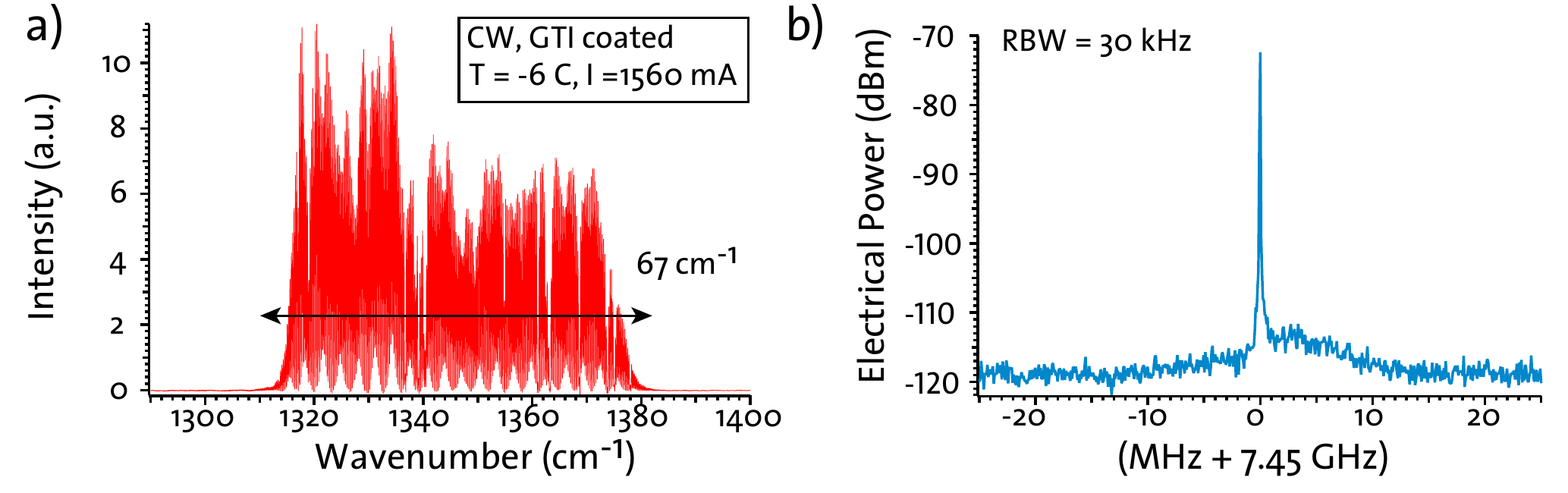}
	\caption{
		High-performance QCL-combs.
		\textbf{a} Optical spectrum of a high performance QCL-comb (6.0 mm long, GTI mirror on the back facet introducing negative dispersion) acquired at T = -6 $^{\circ}$C, I = 1560 mA. The power-per-mode distribution shows a normalized standard deviation of 31 \%.
		\textbf{b} RF spectrum measured at the same value of current than in Fig.\,\ref{fig:Very_High_perf_QCL_comb}\textbf{a}, acquired with a spectrum analyzer (span = 50 MHz, RBW = 30 kHz, acquisition time = 20 ms). The RF spectrum shows a narrow beatnote, characteristic of comb operation, together with a pedestal observed at a level 40 dB lower then the carrier. The signal-to-noise ratio of the RF beatnote is more then 40 dB. 	
	}
	\label{fig:Very_High_perf_QCL_comb}
\end{figure}

The performances of a QCL-comb coated with a GDD mirror introducing negative GDD ($-6814$ fs$^{2}$ introduced at 1258 cm$^{-1}$) is shown in Fig.\,\ref{fig:High_perf_QCL_comb}. We use the same characterization set-up as already described in Fig.\,\ref{fig:QCL_comb_starting_point}\textbf{a}, where optical spectra and RF spectra can be acquired simultaneously. Fig.\,\ref{fig:High_perf_QCL_comb}\textbf{a} shows the power-current-voltage measured in CW operation for a GTI-coated QCL-comb with negative dispersion. As GTI mirrors also act as HR coatings, we observe a decrease of the threshold current as well as an increase of the slope-efficiency. The device emits 142 mW at T = -20 $^{\circ}$C in CW operation. We observe in the RF spectra (see Fig.\,\ref{fig:High_perf_QCL_comb}\textbf{b}) that the beatnotes generated at the comb repetition frequency are extremely narrow (FWHM $<500$ Hz) for all the different values of current. Therefore, the comb regime -- which was present over a small range of the dynamical range of the QCL-comb without GTI mirror -- is now observed over a large dynamic range of the QCL-comb. More importantly, we observe that the device operates in the comb regime until the laser roll-over and that no high-phase noise regime is observed. Finally, as observed in the optical spectra  shown in Fig.\,\ref{fig:High_perf_QCL_comb}\textbf{c}, the power-per-mode distribution is more homogeneous when compared to the QCL-comb with no dispersion compensation (see Fig.\,\ref{fig:QCL_comb_starting_point}\textbf{c}). These findings were observed in several similar devices (same length, same laser fabrication process) which where coated with the same GTI mirror (see Appendix~\ref{App:add_QCL_comb_GTI}). 

High performances QCL-combs are obtained when compensating the dispersion of a 6 mm long device with a GTI mirror introducing negative dispersion ($-6814$ fs$^{2}$ introduced at 1258 cm$^{-1}$). Fig.\,\ref{fig:Very_High_perf_QCL_comb}\textbf{a} shows the optical spectrum of such QCL-comb, acquired when the laser is close to roll-over. The RF spectrum is also measured and is shown in Fig.\,\ref{fig:Very_High_perf_QCL_comb}\textbf{b}. The power-per-mode distribution on the optical spectrum shows a normalized standard deviation of 31 \%. More importantly, the RF spectrum shows that the laser is operating in a comb regime, characterized by a single and narrow RF beatnote (FWHM $< 30$ kHz). Even though we observe a broad pedestal on the RF spectrum, this pedestal is observed at a level 40 dB lower then the carrier. Moreover, the RF beatnote signal-to-noise ratio (40 dB for the beatnote shown in Fig.\,\ref{fig:Very_High_perf_QCL_comb}\textbf{b}) is significantly higher then the ones observed for QCL-combs with a standard HR coating (see Fig.\,\ref{fig:QCL_comb_starting_point}\textbf{c}). Again, no high-phase noise regime was observed on this device.  

\subsection{Discussion and conclusion}

Our experimental results show that QCL-combs can be precisely designed in order to achieve high-performance MIR semiconductor based frequency combs. A detailed experimental analysis of the dispersion introduced on QCL-combs is performed and the concept of an integrated GTI mirror for QCL-combs is introduced. These GTI mirrors were designed in order to introduce positive or negative GDD. Improved designs of GTI mirrors based on different materials (Ge, YF3) could also be implemented in case higher values of dispersion compensation are needed. By fine characterization of the dispersion introduced by the GTI, QCL-combs operating at negative GDD values were obtained. This led to QCL-combs showing a comb regime spanning over a wide current range, and no signature of high-phase noise regime was observed. Moreover, the power-per-mode distribution on these QCL-combs is more homogeneous compared to previously designed QCL-combs. These devices are ideal for systems using QCL-combs for spectroscopy applications, where a highly inhomogeneous power distribution along the comb modes is detrimental for high accuracy spectroscopy, as important values of signal-to-noise ratio are needed over the entire spectrum~\cite{Villares:2014}.

In conclusion, we have demonstrated a high-performance MIR QCL-comb obtained by dispersion compensation. By operating in the negative dispersion regime, the QCL-comb perfomances were dramatically improved. We achieved high power QCL-combs ($\simeq$150 mW) spanning over $\simeq$ 70 cm$^{-1}$, where the comb operation regime is extended over a wide current range and where no signature of the high-phase noise regime is observed. The spectral coverage of the QCL-combs is only limited by the bandwidth of the gain medium. Therefore, by using GTI mirrors to compensate the dispersion of multi-stack QCL designs with broader spectrum, QCL-combs as broad as 300 cm$^{-1}$ could be in principle fabricated. For compensating the dispersion in a wider range, GTI mirrors terminated with dielectric HR coatings could be realized or double GTI designs could also be implemented~\cite{golubovic2000double}. Conversely, the comb structure changes dramatically when operating in the negative dispersion region, as shown by the increase of the comb operation regime and also by the modification of the power distribution spectrum. This is a signature that the control of the dispersion can induce a change in the phase distribution between the comb modes. By measuring the relative phases of the comb modes as well as by measuring an ultra-short temporal profile of the laser intensity, by using a frequency-resolved optical gating technique~\cite{trebino1997measuring} or a ultrafast temporal magnifier~\cite{foster2009ultrafast,salem2009high} -- as recently done on the field of Kerr combs~\cite{herr2014temporal,del2015phase} and also for QCL-combs~\cite{burghoff2015evaluating} -- the structure of a QCL-comb can be fully understood. The control of the comb phases could potentially lead to the creation of QCL-combs operating in comb states not observed to date.         


\section{Methods}\label{sec:methods}

\subsection{QCL-comb characterization}
In order to characterize the performance of QCL-combs, the RF spectrum containing the comb repetition frequency and the optical spectrum have to be acquired simultaneously. The QCL-combs are driven with low noise current drivers (Wavelength electronics QCL500 OEM or QCL2000 LAB) with a specified average current noise density of 2 nA/$\sqrt{\textrm{Hz}}$. The temperature fluctuations of the lasers were also reduced to less than 10 mK by using a low thermal drift temperature controller (wavelength electronics PTC10K-CH) with a 50 k$\Omega$ thermistor. The comb repetition frequency is measured through the RF port of a bias-tee inserted between the current driver and the QCL-comb (c.f. Fig.\,\ref{fig:QCL_comb_starting_point}\textbf{a})~\cite{hugi2012mid,barbieri2011coherent}. A RF spectrum analyzer (Rhode \& Schwarz FSU50) is used to acquire the RF spectrum. 
After collimation by a high-numerical aperture (0.86) aspheric lens, the QCL-comb output is sent to a Fourier Transform Infrared Spectrometer (FTIR, 0.12 cm$^{-1}$ resolution) in order to acquire the optical spectrum (c.f. Fig.\,\ref{fig:QCL_comb_starting_point}\textbf{a}).

\subsection{GTI mirrors for dispersion compensation}
The GTI mirrors designed for our QCL-combs consist of different layers of dieletric materials and a final layer of gold. Ideally, the layers of the GTI have to be transparent or at least introduce negligible absorption. We therefore used a combination of $\textrm{Al}_{2}\textrm{O}_{3}$ and $\textrm{SiO}_{2}$, as they are relatively transparent in the spectral region of the QCL-combs designed for this study (see see Appendix~\ref{App:disp_char}). The operation of the GTI mirror is done such that either a minimum or maximum of the GDD introduced by the mirror lies in the comb spectral region, in order to introduce negative or positive dispersion, respectively. The design is performed numerically by using a simulation tool computing the complex reflectivity of the coating, based on a transfer matrix formalism. However, the starting point of the design is given by the analytical expression of the GDD introduced by a GTI made with a perfect transparent material
\begin{align*}
\textrm{GDD}_{\textrm{GTI}} = - \frac{2\tau_{0}^{2}(1-R_{t})\sqrt{R_{t}}\sin \omega\tau_{0}}{(1 + R_{t} - 2\sqrt{R_{t}}\cos \omega\tau_{0})^{2}}
\end{align*}
where $\tau_{0}$ is the cavity roundtrip time, $R_{t}$ is the reflection coefficient of the low reflectivity facet of the GTI and $\omega$ is the optical frequency. We used the transfer matrix simulation tool only to improve and refine this initial design, in which the layered structure and the absorption of the different layers were not considered. 
The precision of the individual layer thicknesses is of critical importance for the correction of the dispersion introduced by the GTI mirror. To satisfy the required precision, calibration runs were conducted before every GTI evaporation to ensure variations of the thicknesses on the order of 3\% percent.


\section*{End Notes}

\section*{Acknowledgements}
We thank Dr. Andreas Hugi for fruitful discussions. This work was financially supported by the Swiss National Science Foundation (SNF200020 - 152962) and by the ETH Pioneer Fellowship programme.

\subsection*{Author Contributions}
J.W. designed the QCL active region. M.B grew the QC structure. M.J.S. and G.V fabricated the QCL-combs. G.V. designed the GTI mirrors and S.R. fabricated them. G.V. and J.F. developed the algorithm for retrieving the GDD. D.K. and G.V. mounted the devices and carried out the measurements. G.V. wrote the paper and made the figures. G.V. and J.F. joined the discussion and provided comments. All the work has been done under J.F. supervision.

\subsection*{Competing financial interests}
The authors declare no competing financial interests.

\appendix 
\section{Dispersion characterization of QCL-combs} \label{App:disp_char}
In this section, we describe the procedure to measure the dispersion introduced by GTI mirrors as well as the dispersion of a QCL-comb. We first describe the entire procedure for the case of GTI mirrors. The method employed for the dispersion measurement of  QCL-combs being very similar, only the differences will be detailed.

\begin{figure*}[t!]
	\centering
	\includegraphics[width=0.6\textwidth]{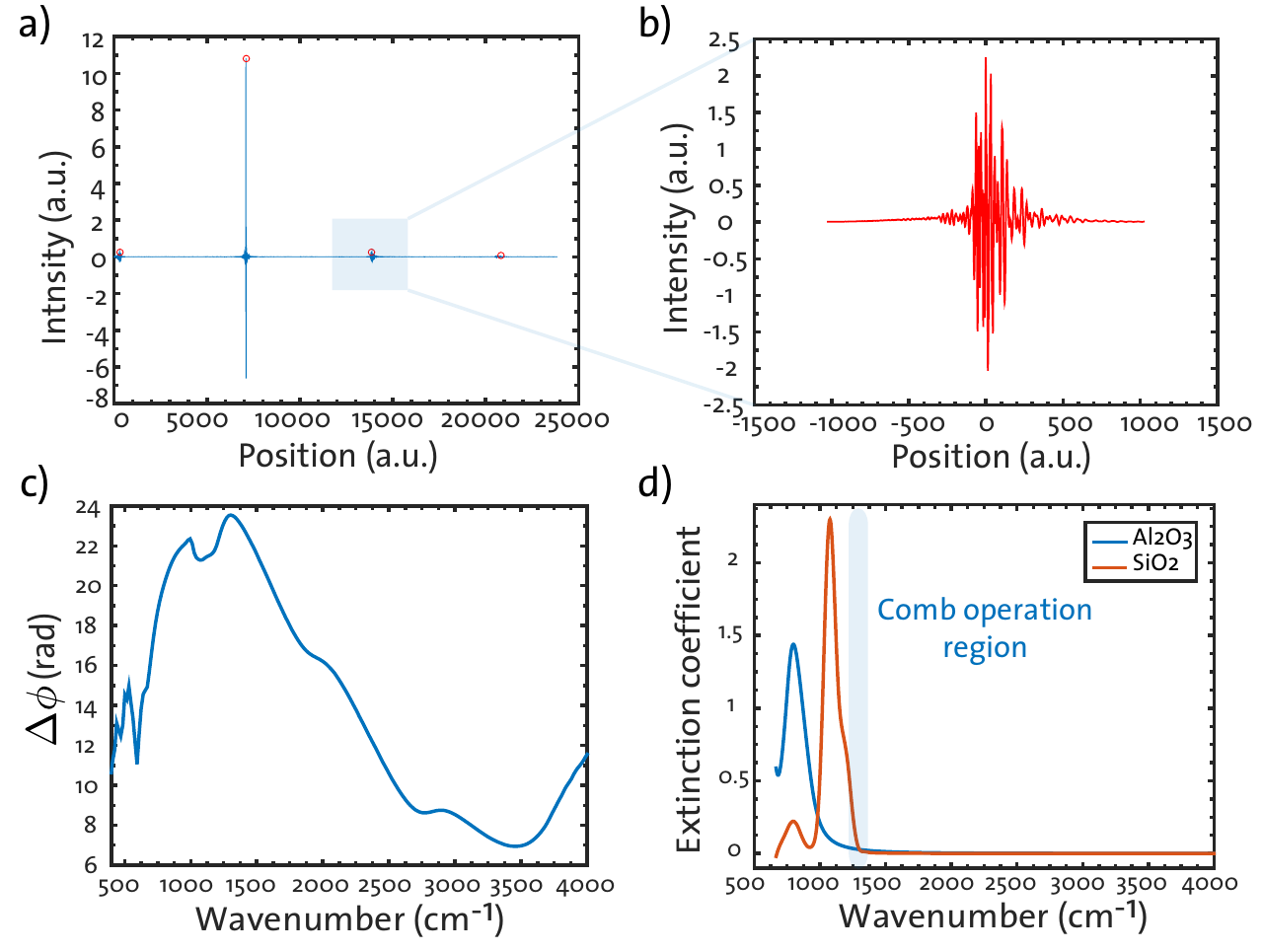}
	\caption{
		Method for dispersion characterization of GTI mirrors as well as QCL-combs.
		\textbf{a} Interferogram generated by a FTIR when measuring the reflection of a GTI mirror.  The red dots correspond to the different satellites observed on the interferogram. The resolution is set to 1.7 cm$^{-1}$, enough to be able to observe up to the second satellite created on the interferogram.
		\textbf{b} Zoom on the first satellite, where the dispersion information is contained. The number of points of this isolated interferogram sets the frequency resolution of the GDD measurement as well as the GDD accuracy.
		\textbf{c} Relative phase spectrum extracted from the first satellite of the interferogram.
		\textbf{d} Extinction coefficient of the dielectric materials ($\textrm{Al}_{2}\textrm{O}_{3}$ and $\textrm{SiO}_{2}$) used in GTI mirrors designed for QCL-combs. The comb spectral region is also highlighted.
	}
	\label{fig:GTI_GDD_method}
\end{figure*}

For each evaporation of a GTI mirror, a InP substrate (320 $\mu$m thick) is placed on the evaporation chamber and is used as a reference sample. This reference sample is used to characterize the dispersion introduced by the GTI mirror. The characterization is done by placing the device under test (DUT) after the beam-splitter on a FTIR, where the beams are recombined, as shown schematicaly on Fig.\,2\textbf{c} of the main text. An interferogram generated by the reflection upon the GTI mirror is then acquired. A typical interferogram is shown in Fig.\,\ref{fig:GTI_GDD_method}\textbf{a}, where we observe an intense center burst, corresponding to the position of the moving mirror where both arms are introducing the same optical delay. More importantly, several small satellites are also observed, indicated by the red dots on Fig.\,\ref{fig:GTI_GDD_method}\textbf{a}. As explained schematically on Fig.\,2\textbf{b} of the main text, these different satellites correspond to the case where light experience multiple roundtrips on the GTI mirror. The first and most intense satelite (position  14000 in Fig.\,\ref{fig:GTI_GDD_method}\textbf{a}) corresponds to the case of a single roundtrip and contains the information concerning the dispersion introduced by the GTI.

In order to calculate the dispersion introduced by the GTI, the first satellite is numerically isolated and apodized, as shown in Fig.\,\ref{fig:GTI_GDD_method}\textbf{b}. The length of this interferogram has to be carfully chosen, as it determines the spectral resolution as well as the accuracy of the GDD measurement~\cite{gosteva2005noise}. After apodization, we perform a fourier transform (Fast Fourier Transform algorithm) and compute the phase spectrum, which is shown in Fig.\,\ref{fig:GTI_GDD_method}\textbf{c}. This phase corresponds to the accumulated phase when light experiences a roundtrip inside the GTI mirror, as shown schematically in Fig.\,2\textbf{b} of the main text. Finally, as the group delay dispersion (GDD) is defined as

\begin{align*}
\Delta\phi(\omega) = \underbrace{\Delta\phi_{0}}_{\textrm{absolute phase at }\omega_{0}} + (\omega-\omega_{0})\underbrace{\Bigg(\frac{d\Delta\phi}{d\omega}\Bigg)_{\omega_{0}}}_{\textrm{Group Delay}} + \frac{1}{2}(\omega-\omega_{0})^{2}\underbrace{\Bigg(\frac{d^{2}\Delta\phi}{d\omega^{2}}\Bigg)_{\omega_{0}}}_{\textrm{GDD}}
\end{align*}

the GDD can be obtained by computing the second derivative of the relative phase $\Delta\phi$ according to $\omega$. The measured GDD is displayed on Fig.\,2\textbf{c} of the main text.

We report now the different designs of GTI mirrors that were used to introduce positive/negative values of dispersion, as described in the main text. Table~\ref{tab:GTI_composition} shows the detailed structure of the different designs of GTI mirrors. The GTI mirrors are composed of different layers of $\textrm{Al}_{2}\textrm{O}_{3}$ and $\textrm{SiO}_{2}$ and are always terminated with a thin layer of gold. As discussed in the main text, a discrepancy between the simulated values of the GDD introduced by a GTI mirro and the measured values of the GDD is observed around 1270 cm$^{-1}$. This is explained by the absorption introduced by $\textrm{SiO}_{2}$. Fig.\,\ref{fig:GTI_GDD_method}\textbf{d} shows the extinction coefficient of both $\textrm{Al}_{2}\textrm{O}_{3}$ and $\textrm{SiO}_{2}$ as well the spectral region were the comb is operating. Although $\textrm{Al}_{2}\textrm{O}_{3}$ can be assumed perfectly transparent in the comb operation region, $\textrm{SiO}_{2}$ is slightly absorbing in this spectral region. This small absorption introduces an additional term to the relative phase $\Delta\phi$ introduced by the GTI, which directly translates into an additional source of dispersion. By using this additional source of GDD, we were able to introduce around -7000 fs$^{2}$ at 1270 cm$^{-1}$, value that was not possible to achieve with a totally transparent GTI mirror with the same thickness. 

\begin{table}[h!]
	\begin{center}
		\begin{tabular}{|p{3cm}|p{3cm}||p{3cm}|p{3cm}| }
			\hline
			\multicolumn{2}{|c||}{GTI structure for positive GDD} 
			& \multicolumn{2}{|c|}{GTI structure for negative GDD}\\
			\hline
			Material & Thickness & Material & Thickness\\
			\hline
			$\textrm{Al}_{2}\textrm{O}_{3}$ & 250 nm& $\textrm{Al}_{2}\textrm{O}_{3}$ & 511 nm \\
			\hline
			$\textrm{SiO}_{2}$ & 250 nm  & $\textrm{SiO}_{2}$&102 nm \\
			\hline
			$\textrm{Al}_{2}\textrm{O}_{3}$ & 250 nm  & $\textrm{Al}_{2}\textrm{O}_{3}$ & 511 nm \\
			\hline
			$\textrm{SiO}_{2}$& 250 nm  & $\textrm{SiO}_{2}$& 102 nm \\
			\hline
			$\textrm{Al}_{2}\textrm{O}_{3}$& 250 nm  & $\textrm{Al}_{2}\textrm{O}_{3}$ & 511 nm \\
			\hline
			$\textrm{SiO}_{2}$ & 250 nm  & $\textrm{SiO}_{2}$& 102 nm \\
			\hline
			$\textrm{Al}_{2}\textrm{O}_{3}$ &250 nm  & $\textrm{Al}_{2}\textrm{O}_{3}$ &511 nm \\
			\hline
			$\textrm{SiO}_{2}$& 250 nm  & $\textrm{SiO}_{2}$& 102 nm \\
			\hline
			$\textrm{Al}_{2}\textrm{O}_{3}$& 250 nm  & $\textrm{Al}_{2}\textrm{O}_{3}$ & 511 nm \\
			\hline
			$\textrm{SiO}_{2}$& 250 nm  & $\textrm{SiO}_{2}$ & 102 nm \\
			\hline
			$\textrm{Al}_{2}\textrm{O}_{3}$& 150 nm  & $\textrm{Al}_{2}\textrm{O}_{3}$ & 154 nm \\
			\hline
			Au & 150 nm  & Au & 154 nm \\
			\hline
		\end{tabular}
		\caption{Detailed structure of GTI mirrors developed for dispersion compensation of QCL-combs. The first evaporated layer correspond to the first layer of the table.}
		\label{tab:GTI_composition}
	\end{center}
\end{table}

Finally, we described the method used for characterizing the dispersion of MIR QCL-combs. The QCL-comb is driven under threshold and is aligned to a FTIR, which is used to acquire an interferogram. The first satellite observed in the interferogram corresponds to the relative phase between the light being directly emitted by one facet and the light experiencing a roundtrip into the cavity and subsequently emitted by this same facet ~\cite{hofstetter1999measurement}. Therefore, the information concerning the dispersion of a QCL-comb is contained into this part of the interferogram. By applying the same numerical method discribed for the analysis of the dispersion introduced by GTI mirrors, the phase spectrum as well as the GDD of a QCL-comb can be computed. The GDD of a QCL-comb coated with a GTI mirror is shown in Fig.3 \textbf{c} of the main text.

\section{Additional QCL-comb with negative GTI mirror} \label{App:add_QCL_comb_GTI}

\begin{figure*}[t!]
	\centering
	\includegraphics[width=0.6\textwidth]{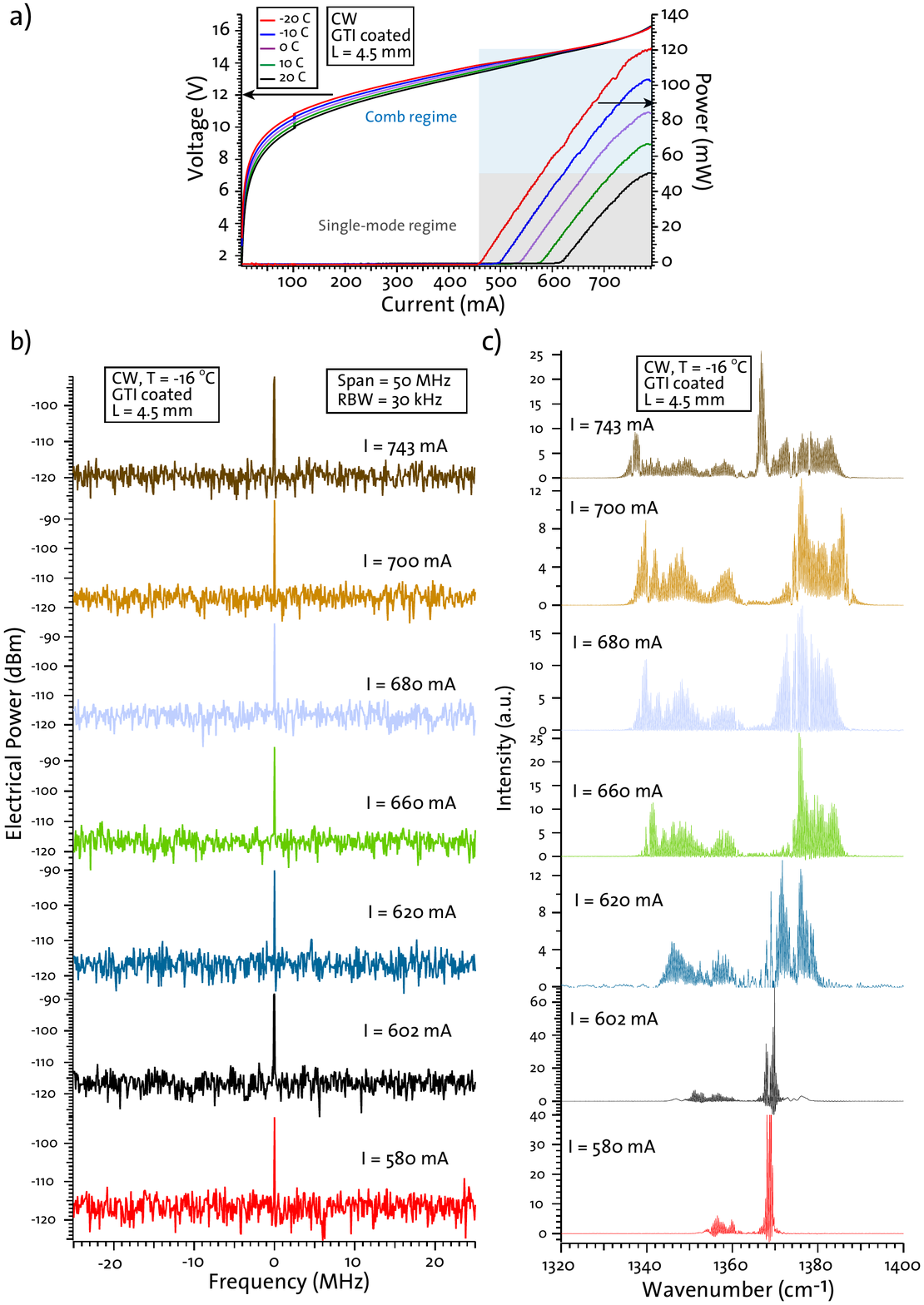}
	\caption{
		Dispersion compensated QCL-combs (additional data).
		\textbf{a} Power-current-voltage of a QCL-comb (4.5 mm long, episide-down mounted on AlN submount) coated with a GTI mirror introducing negative dispersion. The measurements are done in CW operation at different temperatures. Single-mode and Comb regimes are highlighted are highlighted. 
		\textbf{b} Electrical RF spectra acquired at T = -16 $^{\circ}$C at different values of current, measured with a spectrum analyser (span = 50 MHz, RBW = 300 kHz, acquisition time = 20 ms). The RF spectra are centered at 9.814 GHz, corresponding to the RF beatnote created by a 4.5 mm long device. The measured RF spectra show single and narrow beatnotes (FWHM $<$ 30 kHz). No high-phase noise regime is observed. 
		\textbf{c} Optical spectra acquired at T = -16 $^{\circ}$C at the same values of current as in Fig.\,\ref{fig:additional_QCL_comb_GTI}\textbf{b} and acquired with a FTIR (0.12 cm$^{-1}$ resolution). The QCL-comb spectrum is centered at 1360 cm$^{-1}$ and spans over 50 cm$^{-1}$ in the comb regime. 
	}
	\label{fig:additional_QCL_comb_GTI}
\end{figure*}

In this section, we show additional data concerning QCL-combs coated with GTI mirrors introducing negative dispersion. We fabricated several devices with the same length from the same laser fabrication process. The same design of GTI mirror introducing negative dispersion was used for all the evaporations. The device shown in this section is similar to the device shown on section C of the main text (same process, cleaved at the same time) but differs by the fact that the GTI mirror was coated on a different evaporation run. The device operates at room-temperature emitting $>100$ mW of output power in CW operation (see Fig.\,\ref{fig:additional_QCL_comb_GTI}\textbf{a}). No high-phase noise regime is observed for this device, as observed for the device described on the section C of the main text. The device shows single narrow RF beatnotes over the entire current range where multimode operation is observed (see Fig.\,\ref{fig:additional_QCL_comb_GTI}\textbf{a} for the RF spectra and Fig.\,\ref{fig:additional_QCL_comb_GTI}\textbf{b} for the optical spectra as a function of current at a fixed temperature), characteristic of comb regime operation.   

\newpage
\newpage

\bibliographystyle{apsrev}
\bibliography{MyLibrary}
\newpage
\end{document}